\documentstyle[aps,twocolumn,epsf,psfig]{revtex}

\begin{document}

\widetext

\title {Generation of short and long range temporal correlated noises}

\author{Aldo H. Romero~\cite{bline}\\
Department of Chemistry and Biochemistry 0340\\
and Department of Physics\\
University of California, San Diego\\
La Jolla, California 92093-0340\\
\and
Jose M. Sancho\\
Depatament d'Estructura i Constituents de la Mat\'eria,\\
Av. Diagonal 647, 08028 Barcelona. Spain.
}

\date{\today} 

\maketitle

\begin{abstract}

We present the implementation of an algorithm to generate Gaussian
random noises with prescribed time correlations that can be either
long or short ranged. Examples of Langevin dynamics with short and
long range noises are presented and discussed.

\end{abstract}

\narrowtext

\section{Introduction}

In stochastic process simulations, the generation of noise with some 
specific statistical properties has been of great importance.  In most 
cases, the required noise is Gaussian and white (delta correlated); in 
many other cases, as in noises of real systems, a specific correlation 
is needed and an appropriate noise generation has to be implemented 
which can be carried out in standard computer facilities. Despite the
extensive work in developing algorithms to generate noises with 
particular correlations, there is still a lag concerning those noises 
with any given temporal, spatial or spatiotemporal correlation.

Some algorithms have been proposed in the last few years for noises 
which have been proved to obey a linear Langevin equation with a 
linear, Gaussian, white, noise term.  A formal integration of this 
linear equation is enough to generate a random process with a very 
particular time correlation.  Examples of this approach are the so
called Ornstein-Uhlenbeck (O-U) and Wiener (W) processes.  The physical 
meaning of the O-U process is the velocity of a Brownian particle under 
the influence of friction and immersed in a heath bath. W-process is the 
archetype for the dynamical behavior of change in position of a free 
Brownian particle in the high friction limit.  The correlations of these 
two processes are well known and can be obtained 
analytically~\cite{gardiner} and numerically~\cite{nises} 
by standard methods of stochastic processes.

Nevertheless, quite often in numerical stochastic simulations, we are 
faced with processes characterized by a Gaussian noise with a specific 
correlation and an unknown Langevin-like equation dynamics.  Therefore, 
a more general algorithm that only depends on the knowledge of the 
temporal correlation is necessary \cite{Barabasi,makse96}.

The purpose of this work is to present the implementation of an algorithm 
which allows to simulate Gaussian noises with almost any given temporal 
correlation function.  The only requirement for the algorithm to work is 
that the Fourier transform of the temporal correlation function should be 
known\cite{makse96}.  In cases where the Fourier transform of the required 
temporal correlation function can not be properly defined, the algorithm 
can be used with a suitable cutoff.

In Section II, algorithms based in Langevin equations are revised and the 
new algorithm of noise generation is presented in practical terms. The 
algorithm has the property of simulating different correlation regimes, 
from short to long ranged.  Section III considers different applications 
for which the dynamics of the system are very sensitive to the noise 
correlation.  A summary of results and some conclusions are presented in 
the last section.

\section{Algorithms to Generate Gaussian Noises}

\subsection{The Linear Langevin Equation Method}

For the sake of comparison with the method we want to present here, let 
us revise briefly the main points of the generation of a Gaussian noise 
which follows a linear Langevin equation in terms of a Gaussian white 
noise.  The stochastic discretized trajectories are generated by formally 
integrating the corresponding Langevin equation and using a set of 
Gaussian random numbers for a given realization.  The statistical 
properties are calculated from many realizations of these trajectories 
(``an ensemble''). In the case of Gaussian noise, only two moments are 
necessary to characterize the statistical properties of the random process.
Due to the linear character of the Langevin equation the Gaussian property 
of the white noise is transmitted to the generated noise.

The simplest case is the W-process, which follows the Langevin equation
\begin{equation}
{\dot \eta} \ = \ \xi(t)~,
\label{W}
\end{equation}
where $\xi(t)$ is a Gaussian white noise of zero mean and delta correlation:
\begin{equation}
\langle \xi( t) \xi(t') \rangle \ = \ 2 \epsilon \delta(t-t')~.
\end{equation}
Here, $\epsilon$ is the noise intensity.

The statistical properties are easily evaluated \cite{gardiner} and a 
trajectory for this type of noise can be obtained by formally integrating 
Eq.(\ref{W}) over time:
\begin{equation}
\eta(t+ \Delta t) \ = \ \eta(t) \ + \ \int^{t + \Delta t}_t \xi(t') dt' \ = 
\ \eta(t) \ + \ X(t)~, 
\label{W-alg}
\end{equation}
where the noisy term $X(t)$, is constructed from
\begin{equation}
X(t) \ = \ \sqrt { 2 \epsilon \Delta t }~~\alpha(t)~.
\end{equation}
Here, $\alpha(t)$ is a set of Gaussian independent random numbers of zero 
mean and variance equal to unity, obtained from any reliable Gaussian 
random generator~\cite{Press,Toral}.

As a second example, we consider the O-U process.
This process contains a temporal memory and obeys the Langevin equation
\begin{equation}
{\dot \eta} \ = \ - \ \frac{ \eta}{\tau} \ + \ \frac{\xi(t)}{\tau}~,
\label{OU}
\end{equation}
where $\tau$ is the characteristic time memory. As in the former case, a 
formal integration gives:
\begin{equation}
\eta(t+ \Delta t) \ = \ \eta(t) e^{- \frac{\Delta t}{\tau}} \ + \
\frac{\epsilon}{\tau} \int^{t + \Delta t}_t  e^{- \frac{ t-t'}{\tau}}~
\xi(t')~dt'~.
\end{equation}
Studying the statistical properties of the noisy term, the algorithm 
reads\cite{laure},
\begin{equation}
\eta(t+ \Delta t) \ = \ \eta(t) e^{- \frac{\Delta t}{\tau}} \ + \ 
\sqrt{ \frac{ \epsilon}{\tau} ( 1 - e^{- \frac{ 2 \Delta t }{\tau}})} 
\ \alpha(t)~.
\label{OU-alg}
\end{equation}  

In this way, a time stochastic trajectory is generated step by step as 
in the former example.  Moreover, in the algorithm we introduce below,  
the whole trajectory is constructed in a single calculation step.

\subsection{Spectral method}

As we noted in the introduction, this method starts from the knowledge 
of the time correlation function.  Probably the basis of this algorithm 
has been rediscovered and implemented several times, but now, due to the 
actual high speed and large storage space in computers, the algorithm can 
be implemented rather easily. Here we follow the main ideas of 
Ref.~\cite{Barabasi,makse96} and first introduce some steps that simplify the 
calculations.

We want to generate a Gaussian, random correlated, noise, $\eta(t)$, 
whose correlation function $\gamma (t)$, is defined by:
\begin{equation}
\langle \eta (t) \eta (t') \rangle \ = \ \gamma (|t - t'|)~,
\label{corret}
\end{equation}
and its Fourier transform, 
\begin{equation}
\gamma(\omega) \ = \ \int e^{-i \omega t } \gamma(t) dt~.
\label{correw}
\end{equation}
is known (to some extent).
In the $\omega$-Fourier space, this correlation reads:
\begin{equation}
\langle \eta(\omega) \eta (\omega') \rangle \ = \ 
2 \pi \delta( \omega + \omega')~\gamma(\omega)~,
\end{equation}
where $\eta(\omega)$ is the Fourier transform of $\eta(t)$. 

With this initial information in mind, the algorithm could be summarized 
as follows.  First we discretize the time in $N=2^n$ intervals of mesh 
size $\Delta t$, and note that this time interval has be much smaller 
than any other characteristic time of the system for our method to work.
Every one of these intervals will be denoted by a Roman-index in real 
space (time) and by a Greek-index in Fourier space (frequency). 

\begin{figure}
\centerline{\psfig{file=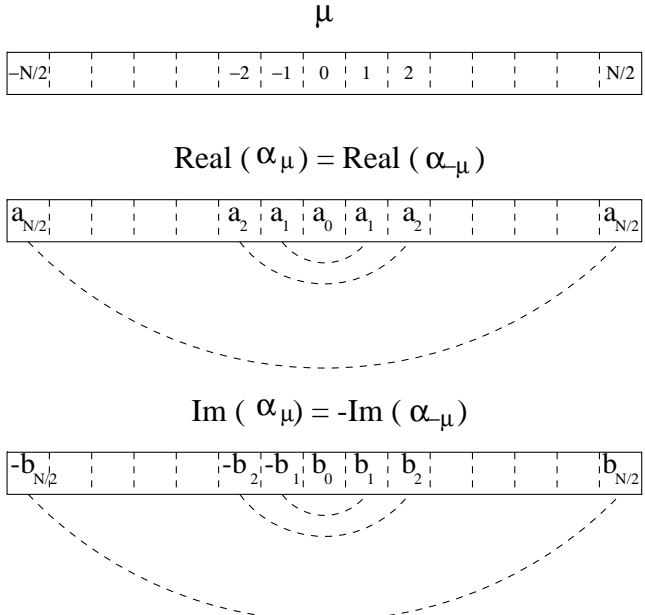,width=8.5cm}}
\caption
{\em 
Diagram representing the construction of a discrete delta anticorrelated
noise in Fourier space. Indexes $a_i$ and $b_i$ represents
Gaussian random numbers as explained in the text.}
\label{ksym}
\end{figure}

In the discrete Fourier space, the noise has a correlation given by:
\begin{equation}
\langle \eta(\omega_\mu) \eta (\omega'_{\mu'}) \rangle \ = \ 
N \Delta t~~ \delta_{\mu + \mu' , 0} \ \gamma ( \omega_{\mu})~.
\end{equation}
Now, $\eta(\omega_\mu)$ can be constructed from
\begin{eqnarray}
\eta(\omega_\mu) \ = \ \sqrt{ N \Delta t \gamma(\omega_\mu)} \ 
\alpha(\omega_\mu)~, \qquad \qquad \qquad \qquad 
\nonumber
\\
\mu \ = \ 0...N, \qquad \eta(\omega_0) \ = \ \eta(\omega_N)~, 
\qquad \omega_\mu \ = \ \frac{2 \pi \mu}{N \Delta t}~,
\label{noisew}
\end{eqnarray}
where $\alpha(\omega_\mu) \equiv \alpha_{\mu}$ are Gaussian random numbers 
with zero mean and correlation,
\begin{equation}
\langle \alpha_\mu \alpha_\nu \rangle \ = \ \delta_{ \mu +\nu,0}~.
\label{anticor}
\end{equation}
This type of delta-anticorrelated noise can be generated rather easily
if the symmetry properties of real periodic series in the Fourier space 
($\alpha_i$) are used \cite{jordiPRE}.  Thus, we avoid the extra work 
involved in Fourier transforming real random numbers\cite{makse96}.  
Consider for example a system of size $N$, the Fourier components of a 
periodic series are then related by:
\begin{equation}
\alpha_\mu \ = \ \alpha_{\mu+pN} \qquad \qquad \alpha_\mu \ = \ 
\alpha^*_{-\mu}~,
\end{equation}
where $p$ is an integer number. Since $\alpha_{\mu = 0} $is real and 
$Im(\alpha_{\mu > 0}) \ = \ - Im(\alpha_{\mu < 0})$, as can be seen in 
figure~\ref{ksym}, the requested anticorrelated complex random
numbers (\ref{anticor}) can be constructed as
\begin{equation}
\alpha_\mu \ = \ a_\mu \ + \ i b_\mu~, \qquad \qquad b_0 \ = \ 0~,
\end{equation}
where $a_\mu$ and $b_\mu$ represent Gaussian random numbers with zero 
mean and a variance of one half:
\begin{equation}
\langle a_\mu^2 \rangle \ = \ \langle b_\mu^2 \rangle \ = \ \frac{1}{2}~, 
\qquad \qquad \mu \ \neq \ 0~, \qquad \qquad a_0^2 = 1~.
\end{equation}

The discrete inverse transform of $\eta(\omega_\mu)$ is then numerically 
calculated by a Fast Fourier Transform algorithm~\cite{Press}. The result 
is a string of N numbers, $\eta(t_i)$, which by construction, have the 
proposed time correlation~(\ref{corret}).  However, due to the symmetries 
of the Fourier Transform, only $N/2$ of these values are actually 
independent and the remaining numbers are periodically correlated with 
them. 

In order to check the suitability of the procedure, the time correlation
of Equation~(\ref{corret}) is numerically evaluated by an independent 
(non Fourier based) method, namely,
\begin{equation}
\gamma(t_i) \ = \ \Big\langle 
\frac{\sum_{j=0}^{N_{max}} \eta(t_j + i \Delta t) \eta(t_j)}{N_{max} + 1}~
\Big\rangle~. 
\label{numcorre}
\end{equation}
Here, $N_{max}$ is a number smaller than $N/2$ (in the examples that follow, 
it is taken equal to $N/4$).

We now present several examples where this approach is implemented. Since 
the number of applications making use of temporal random noises is quite 
large we have chosen applications with rather different correlation 
properties in order to illustrate the power of the method.  The same 
procedure has already been used for spatial correlated noises~
\cite{jordi97,jordi98,Romero98}.

\subsection{Short range correlated noises}

{\bf a.1}. A Gaussian Correlation.

Let us consider a noise with a Gaussian correlation function defined as 
\begin{equation}
\langle \eta (t) \eta (t') \rangle \ = \ \gamma (|t - t'|) \ = \ \frac{ 2 
\epsilon}{\tau \sqrt{2 \pi} } e^{- \frac{|t - t'|^2}{2 \tau^2}}~,
\label{gausscorr}
\end{equation} 
where $\epsilon$ and $\tau$ are the noise intensity and correlation
time, respectively. The correlation is normalized in such a way that
\begin{equation}
\epsilon \ = \ \int_0^\infty \gamma(t) \ dt~.
\end{equation}
Setting $\tau \rightarrow 0 $, the white noise limit is recovered. The
Fourier transform of this Gaussian correlation(\ref{gausscorr}) is given 
by:
\begin{equation}
\gamma (\omega) \ = \ 2 \epsilon e^{ -\frac{ \tau^2 \omega^2}{4}}~.
\label{gausscorrw}
\end{equation}

According to the prescription~(\ref{noisew}), we now generate the discrete 
field $\eta(\omega_\mu)$ as:
\begin{equation}
 \eta(\omega_\mu) \ = \ \left(N \Delta t \  2 \epsilon e^{\frac{ \tau^2}{ 
\Delta t^2}(cos(2 \pi \mu /N) - 1)} \right) ^{1/2} \alpha_\mu ~.
\label{gausspres}
\end{equation} 

\begin{figure}
\centerline{\psfig{file=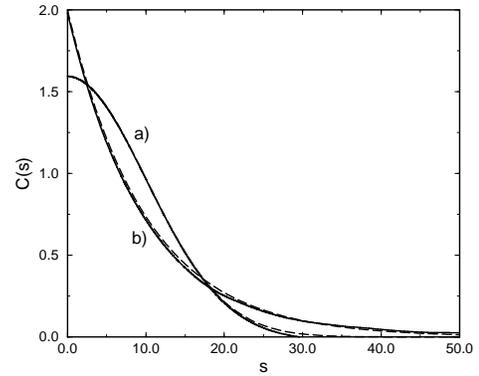,width=7cm}}
\caption
{\em 
Temporal correlation for two short range correlated noises. a) Gaussian
correlation (\protect\ref{gausscorr}). b) Orstein-Uhlenbeck process
(\protect\ref{OUcorr}).
Common parameters: $N=2^{17}, \Delta t = 0.01, \epsilon = 20, \tau = 10$.
Full lines are the simulation data and dashed lines are the corresponding
theoretically expected  results. 
}
\label{fig:corrshort}
\end{figure}

In Fig~\ref{fig:corrshort}, an explicit comparison between the
numerical results and the expected theoretical prediction is presented. 

{\bf a.2}. The Ornstein-Uhlenbeck process.

The Ornstein-Uhlenbeck process simulates the behavior of the velocity of a 
Brownian particle under friction and immersed in a thermal bath. Quite often, 
it is used to represent a real noise with memory, whose intensity is 
$\epsilon$ and whose correlation time (or memory intensity) is $\tau$.
This is a well known Gaussian, Markovian, and stationary noise, which obeys 
a linear Langevin equation (\ref{OU}). Its well known correlation is given 
by:
\begin{equation}
\langle \eta (t) \eta (t') \rangle \ = \ \gamma (|t - t'|) \ = \ \frac 
{ \epsilon }{  \tau } e^{- \frac{|t - t'|}{\tau}}~,
\label{OUcorr}
\end{equation}
with the same normalization used for the previous example.  Since this noise 
obeys a linear Langevin equation, which was not the case for the first 
example, a realization of this noise could be simulated using  the formal 
solution of the stochastic differential equation given by Eq.(\ref{OU-alg}).  
Instead, we present in this subsection the results obtained following our 
spectral method.  The Fourier transform of this particular correlation 
function is:
\begin{equation}
\gamma (\omega) \ = \ \frac {2 \epsilon}{ 1 + ( \tau \omega )^2 } ~.  
\label{OUcorrw}
\end{equation}

According to the prescription (\ref{noisew}), we generate a discrete 
field $\eta(\omega_\mu)$ as:
\begin{equation}
\eta(\omega_\mu) \ = \ \left( \frac {  N \Delta t \ 2\epsilon}{ 1 + ( 
\frac {2 \tau}{\Delta t} sin( \pi \mu / N ))^2 }\right)^{1/2} \alpha_\mu~.
\label{OUpres}
\end{equation}
 
In Fig. \ref{fig:corrshort}, we compare the numerical and theoretical 
results.  The discretization of the $\omega$--variable using either the 
function $\cos(\frac{2 \pi \mu}{N})$ as in Eq.(\ref{gausspres}) or the 
function $\sin(\frac{\pi \mu}{N})$ as in Eq. (\ref{OUpres}) does not make 
any difference.

We have seen that in the short range correlated noises considered here, 
the agreement between the statistical properties of the noise generated 
from the spectral method and the statistical properties required from 
the noise are quite good.

\subsection{Long range correlated noises}

A more dificult noise, in the sense that it can not be obtained from any
known linear Langevin equation, is the one characterized by a power-law 
decaying correlation function,
\begin{equation}
\gamma(|t - t'|) \ \sim \ \frac {\epsilon}{ |t - t'|^\beta}~, \qquad \qquad 
0<\beta< 1.
\label{powercorr}
\end{equation}
This correlation is not well defined in Fourier space (it has a singularity 
at $|t - t'| = 0$).  Therefore, in order to implement the spectral method, 
we start with a guess for $\gamma(\omega_\mu)$, and then we look at the 
dynamics of $\gamma(t)$ in real space. The Fourier values of the noise are 
now discretely generated according to the expression,
\begin{equation}
\eta(\omega_\mu) \ = \ \frac { N \Delta t \ \epsilon}{\left[ \frac {2 \tau}
{\Delta t} \sin( \pi \mu / N ) + \omega_0 \right]^{(\beta-1)/2}} \ 
\alpha_\mu~, 
\end{equation}
where $\omega_0 $ is a predefined cut-off.  In Ref.\cite{makse96} modified 
Bessel functions are used for the correlation in Fourier space but they 
have the same long range decay.  In our case, we assume the following form 
for the correlation function in real space,
\begin{equation}
\gamma(t) \ = \ \frac {A \epsilon}{ \pi \beta (t + t_0)^\beta}, \qquad \qquad 
0 \ < \ \beta \ < \ 1.
\label{corrlong}   
\end{equation}  
where $t_0= \Delta t / \pi $, $\beta$ is the parameter describing the power 
law decay, $\epsilon$ is the noise intensity, and $A$ is a parameter to be 
fitted from a correlation average.

\begin{figure}
\centerline{\psfig{file=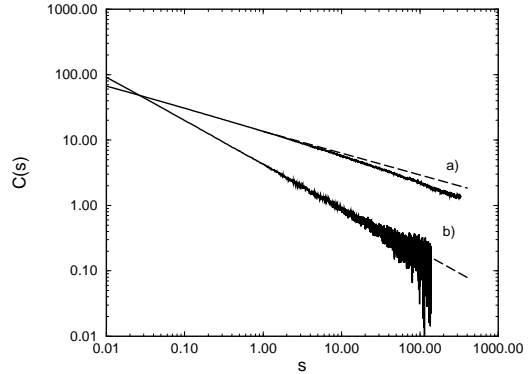,width=7cm}}
\caption{\em Temporal correlation for two long range correlated noises,
with correlation defined by equation (\protect\ref{powercorr}). a)
$\beta = 1/3, \epsilon = 20, A= 0.7095$.
b) $\beta = 2/3, \epsilon = 20, A= 0.4483$.
Common parameters: $N=2 ^{17}, \Delta t = 0.01$.
Full lines are the simulation data and dashed lines are the corresponding
theoretical fitting from equation (\protect\ref{corrlong}).
 }
\label{fig:corrlong}
\end{figure}

Fig. \ref{fig:corrlong}, presents two examples of the power-law Gaussian 
noise with $\beta =1/3$ and $\beta =2/3$.  It can be seen that the numerical 
results are fitted very well by the expected power laws. 

The existence of power law noises has been predicted and discussed for 
quite some time in the literature.  Many examples arise from theoretical 
as well as from experimental studies.  A recent reported case is the 
examination of the temperature fluctuations in climatological data~
\cite{Eva,Eva1}, where a power law functional form is found for the 
correlation of these fluctuations.  In this particular case, the analysis 
is carried over a temperature autocorrelation function as defined by Eq.~(
\ref{numcorre}), and it is found that the distribution of temperature 
fluctuations is well described by a Gaussian distribution with the 
long-ranged decaying auto-correlation function: $C(s) \sim s^{-\gamma}$.  
The exponent found for their data is around $2/3$.  Noises with these 
properties can be generated using our procedure and can be implemented 
in modelling processes similar to the one reported in this climatological
anylisis. 

\begin{figure}
\centerline{\psfig{file=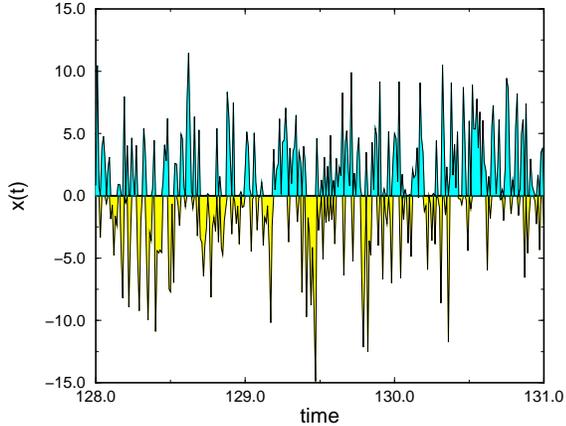,width=8cm}}
\caption{\em
Section of the time series realization for a power law
correlation with $\beta = 2/3$}
\label{fig:smallxt}
\end{figure}

Fig.~\ref{fig:smallxt} shows a realization of a particular noise trajectory 
generated with our method for a  power law noise with $\beta = 2/3$. The 
variations of the data generated with the algorithm look very similar
to figures 1.b, 6.a and 6.b in ref.~\cite{Eva}.  The persistence of the 
series is evident from our simulation where the stochastic trajectory 
appears as ``packets''. 

\section{Applications}

We proceed now to present and discuss two different non-trivial examples 
in non-equilibrium statistical physics where our algorithm can be applied: 
(i) the dispersion process of a Brownian particle and (ii) the decay from 
an unstable state, both cases under the influence of long range noises 
generated through the spectral method.  In the first case the noise is 
additive whereas in the second case it appears multiplicatively.

\subsection{Superdiffusive motion}

In this section we want to focus our attention on the random motion of a 
Brownian particle which obeys the Langevin equation 
\begin{equation}
\frac{dx}{dt} \ = \ \eta(t)~,
\end{equation}
where $\eta$ is a Gaussian noise with a given correlation function.  We 
consider three different cases, a short range and two long range noises.

\begin{figure}
\centerline{\psfig{file=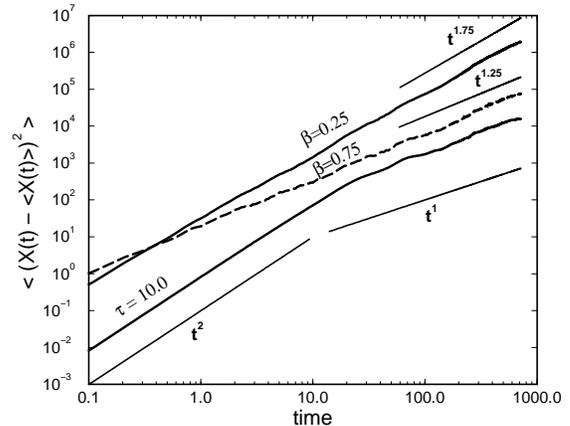,width=8cm}}
\caption{\em
Dispersion as a function of time for three different realizations of the 
noise.  The two upper curves correspond to a power law noise with two 
different exponents, $\beta = 0.25$ and $\beta = 0.75$. The bottom curve 
corresponds to an exponential noise with $\tau = 10.0$.  Straight lines are 
plotted as an eye guide.
}
\label{fig:rms_add}
\end{figure}

The solution for the relative dispersion is well known:
\begin{equation}
\langle \delta x(t)^2 \rangle \ = \ 2 \int_0^t \int_0^{t'} \gamma(s) \ ds \ 
dt'.
\label{dispersion}
\end{equation}
For large times, this expression is either $ \sim t$ (diffusion) for short 
range noises, or $\sim  t^{2 - \beta}$ (super--diffusion) for long range 
noises.  The explicit form of the time dependence of this quantity appears 
in Fig. \ref{fig:rms_add}, where we also show the variance of the three 
different examples in which our algorithm is applied. Two super--diffusive 
cases are clearly seen, with dispersion exponents $\sim 1.75 $ and $1.25$, 
corresponding to noise decay power laws, $\beta = 0.25$ or $0.75$, 
respectively. A short range noise (exponential) is also included for 
comparison.  We see that in this last case the behavior is ballistic 
(deterministic) $ \sim t^2$ for  $t < \tau$, but diffusive for $t \gg \tau$ 
as predicted by Eq.~(\ref{dispersion}).

\subsection{Decay of an unstable state}

The decay of an unstable state is one of many interesting problems which 
appear in nonequilibrium phenomena and nonlinear relaxation process.  The 
switch-on process of a dye laser has been a prototype of a nonequilibrium 
situation in which the influence of several sources of noise have been 
tested.  The dynamics of this system can discriminate the effects of both 
additive or multiplicative noises \cite{roy,pasquale1,pasquale2}.  It has 
been established that additive noise is responsible for the short time 
dynamics but multiplicative noise effects appear in the medium and long 
term dynamics.  In previous studies, white or short range, colored noises 
have been used, but long range noises have never been considered. Here we 
will present several numerical simulation results for the decay of an 
unstable state under the influence of long range multiplicative noises.

In order to simplify the analysis, let us illustrate the situation with the 
following Langevin equation,
\begin{equation}
\frac{dx}{dt} \ = \ a x \ - \ b x^3 \ + \ x \eta(t)~,
\label{multip}
\end{equation}
where $\eta$ is a Gaussian noise whose correlation function is to be 
specified.  For $a$ and $b$ positive parameters, the initial value $x(0)=0$ 
is an unstable state which does need small perturbation to start relaxing 
towards its steady state.  To trigger this process we need either the 
additive noise of Eq. (\ref{multip}) or an initial distribution for $x(0)$.
We have chosen this last assumption and the initial values of $x(0)$ are 
Gaussian distributed with statistical moments $\langle x(0) \rangle = 0 $ 
and $\langle x(0)^2 \rangle = \sigma^2 $.  Due to the symmetries of the 
problem, the mean value is $\langle x(t) \rangle = 0 $, and hence we look 
at the dynamical evolution of the second moment.

To get a precise idea of the short time behavior, it is enough to study 
Eq. (\ref{multip}) in a linear approximation. Formal integration of this
equation gives,
\begin{equation}
x(t)\ = \ x(0) \ exp \left( a t + \int_0^t \eta(t') \ dt' \right)~.
\end{equation}
Using Gaussian properties through the calculation, we obtain for the 
second moment 
\begin{equation}
\langle x^2(t) \rangle \ = \  \sigma^2 \ exp \left( 2at + 4 \Omega(t) \right)~,
\label{2ndmom}
\end{equation}
where $\Omega (t)$ is defined by:
\begin{equation}
\Omega(t) \ = \ \int_0^t \int_0^{t'} \gamma(s) \ ds \ dt'~.
\label{Omega}
\end{equation}

For the deterministic case, $\epsilon = 0$, and the multiplicative white 
noise case, we have: 
\begin{equation}
\langle x^2(t) \rangle \ = \ \sigma^2 \ exp \left( 2at \right)~,
\label{deterministic}
\end{equation}
\begin{equation}
\langle x^2(t) \rangle \ = \ \sigma^2 \ exp \left( 2at +  4 \epsilon t 
\right)~,
\label{whiten}
\end{equation}
respectively.

As in the superdiffusion case, we get for long range noises a power time 
dependence $ \sim t^{2-\beta} $ in the exponential; we therefore expect a 
larger rate for the relaxation of the initial state.

\begin{figure}
\centerline{\psfig{file=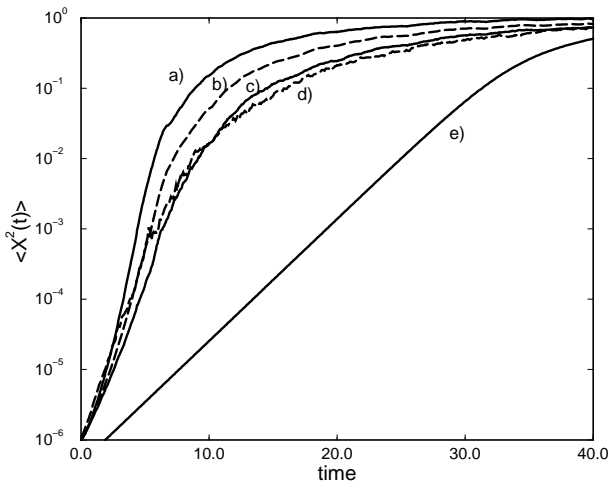,width=8cm}}
\caption{\em
Time evolution of the second moment for different cases.
a) $\beta = 1/3$, b) $\beta = 1/2$, c) $\beta = 2/3$, d) white noise and 
e) deterministic. $ a = b = \epsilon = 0.2 $. }
\label{fig:mnoise}
\end{figure}

In Fig. \ref{fig:mnoise} we see how the decay of the unstable initial state 
is influenced in the different cases discussed. For the values of the 
parameters we have used, Eqs.(\ref{deterministic}) and (\ref{whiten}) predict 
that the multiplicative white noise should decay six times faster than that 
the deterministic case.  For long range correlated noises we expect that the 
smaller the exponent $\beta$ is, the faster it will decay.  Within 
statistical errors, these points are actually seen in the simulation data 
presented in the figure . 
 
It is interesting to note that for intermediate and long times no 
analytical results can be obtained. However, figure~\ref{fig:mnoise}
is telling us that the final steady state also depends on the correlation 
of the noise, taking larger values for smaller $\beta$'s.  This is not an 
easy problem and it would need further theoretical analysis.

\section{Summary and Conclusions}

We have presented a method to generate Gaussian noises with a prescribed time 
correlation function which does not depend on any type of dynamics.  The 
algorithm herein incorporates whithin the same framework the generation of 
long and short ranged noises.  In particular, the need for a good algorithm 
is manifest when long range noises representing physical dynamical process 
with unknown or very complicated dynamical equations are required. In this 
sense, this method can numerically simulate the influence of long and short 
ranged noises in physical systems in a very reliable and controlled way.  
With the prescription given, many problems with long range realizations can 
be simplified and the possibility of some kind of analytical approach to 
specific problems increases (such is the case for the examples we have 
presented). 

The generalization of the algorithm to spaciotemporal noises is straight 
forward for any number of dimensions.  Since by construction, the noise 
generation is finite, the upper time limit of the simulation has to be 
selected in advance, but it does not constitute a serious limitation for 
the algorithm.

\section{Acknowledgment}

The work was supported in part by the U.S. Department of Energy under Grant 
No. DE-FG03-86ER13606, and in part by the Comisi\'on Interministerial de 
Ciencia y Tecnolog\'ia (Spain) Project No. DGICYT PB96-0241.
A.R. thanks K. Lindenberg and A. Sarmiento for support and
helpful comments.

\end{document}